\documentstyle[epsf,epsfig,here,12pt]{article}  
\begin{document}
\newcommand{\fd}{f_{\Bdb}}
\newcommand{\epem}{e^+e^-}
\newcommand{\rhobar}{\overline{\rho}}
\newcommand{\etabar}{\overline{\eta}}
\newcommand{\epsilonk}{\left | \epsilon_K \right |}
\newcommand{\vubovcb}{\left | \frac{V_{ub}}{V_{cb}} \right |}
\newcommand{\vtdovts}{\left | \frac{V_{td}}{V_{ts}} \right |}
\newcommand{\dmd}{\Delta m_d}
\newcommand{\dms}{\Delta m_s}
\newcommand{\pr}{{\rm P.R.}}
\newcommand{\ps}{{\rm ps}}
\newcommand{\Ds}{{\rm D}_s^+}
\newcommand{\Dsm}{{\rm D}_s^-}
\newcommand{\Dp}{{\rm D}^+}
\newcommand{\Dm}{{\rm D}^-}
\newcommand{\Do}{{\rm D}^0}
\newcommand{\Dob}{\overline{{\rm D}}^0}
\newcommand{\piss}{\pi^{\ast \ast}}
\newcommand{\pis}{\pi^{\ast}}
\newcommand{\bbar}{\overline{b}}
\newcommand{\cbar}{\overline{c}}
\newcommand{\Dstar}{{\rm D}^{\ast}}
\newcommand{\Dstars}{{\rm D}^{\ast +}_s}
\newcommand{\Dstaro}{{\rm D}^{\ast 0}}
\newcommand{\Dstarp}{{\rm D}^{\ast +}}
\newcommand{\Dstarstar}{{\rm D}^{\ast \ast}}
\newcommand{\Dstarstarp}{{\rm D}^{\ast \ast +}}
\newcommand{\pistar}{\pi^{\ast}}
\newcommand{\pisstar}{\pi^{\ast \ast}}
\newcommand{\Dbar}{\overline{{\rm D}}}
\newcommand{\Bbar}{\overline{{\rm B}}}
\newcommand{\Bsbar}{\overline{{\rm B}^0_s}}
\newcommand{\Lcbar}{\overline{\Lambda^+_c}}
\newcommand{\nubar}{\overline{\nu_{\ell}}}
\newcommand{\tautaubar}{\tau \overline{\tau}}
\newcommand{\Vcb}{\left | {\rm V}_{cb} \right |}
\newcommand{\Vub}{\left | {\rm V}_{ub} \right |}
\newcommand{\Vtd}{\left | {\rm V}_{td} \right |}
\newcommand{\Vts}{\left | {\rm V}_{ts} \right |}
\newcommand{\fleisher}{\frac{BR({\rm B}^0~(\overline{{\rm B}^0}) \rightarrow \pi^{\pm} {\rm K}^{\mp})}
{BR({\rm B}^{\pm} \rightarrow \pi^{\pm} {\rm K}^0)}}
\newcommand{\bptre}{\rm b^{+}_{3}}
\newcommand{\bp}{\rm b^{+}_{1}}
\newcommand{\bo}{\rm b^0}
\newcommand{\bos}{\rm b^0_s}
\newcommand{\bss}{\rm b^s_s}
\newcommand{\qq}{\rm q \overline{q}}
\newcommand{\cc}{\rm c \overline{c}}
\newcommand{\BsDmX}{{B_{s}^{0}} \rightarrow D \mu X}
\newcommand{\BsDsm}{{B_{s}^{0}} \rightarrow D_{s} \mu X}
\newcommand{\BsDsX}{{B_{s}^{0}} \rightarrow D_{s} X}
\newcommand{\BDsX}{B \rightarrow D_{s} X}
\newcommand{\BDomX}{B \rightarrow D^{0} \mu X}
\newcommand{\BDpmX}{B \rightarrow D^{+} \mu X}
\newcommand{\Dsfmn}{D_{s} \rightarrow \phi \mu \nu}
\newcommand{\Dsfipi}{D_{s} \rightarrow \phi \pi}
\newcommand{\DsfX}{D_{s} \rightarrow \phi X}
\newcommand{\DpfX}{D^{+} \rightarrow \phi X}
\newcommand{\DofX}{D^{0} \rightarrow \phi X}
\newcommand{\DfX}{D \rightarrow \phi X}
\newcommand{\DsD}{B \rightarrow D_{s} D}
\newcommand{\DsmX}{D_{s} \rightarrow \mu X}
\newcommand{\DmX}{D \rightarrow \mu X}
\newcommand{\Zbb}{Z^{0} \rightarrow \rm b \overline{b}}
\newcommand{\Zcc}{Z^{0} \rightarrow \rm c \overline{c}}
\newcommand{\Rbb}{\frac{\Gamma_{Z^0 \rightarrow \rm b \overline{b}}}
{\Gamma_{Z^0 \rightarrow Hadrons}}}
\newcommand{\Rcc}{\frac{\Gamma_{Z^0 \rightarrow \rm c \overline{c}}}
{\Gamma_{Z^0 \rightarrow Hadrons}}}
\newcommand{\bb}{\rm b \overline{b}}
\newcommand{\str}{\rm s \overline{s}}
\newcommand{\Bs}{\rm{B^0_s}}
\newcommand{\Bsb}{\overline{\rm{B^0_s}}}
\newcommand{\Bp}{\rm{B^{+}}}
\newcommand{\Bm}{\rm{B^{-}}}
\newcommand{\Bo}{\rm{B^{0}}}
\newcommand{\Bd}{\rm{B^{0}_{d}}}
\newcommand{\Bdb}{\overline{\rm{B^{0}_{d}}}}
\newcommand{\Lb}{\Lambda^0_b}
\newcommand{\Lbb}{\overline{\Lambda^0_b}}
\newcommand{\Kstar}{\rm{K^{\star 0}}}
\newcommand{\phim}{\rm{\phi}}
\newcommand{\Dsp}{\mbox{D}_s^+}
\newcommand{\Dn}{\mbox{D}^0}
\newcommand{\Dsb}{\overline{\mbox{D}_s}}
\newcommand{\Dnb}{\overline{\mbox{D}^0}}
\newcommand{\Lc}{\Lambda_c}
\newcommand{\Lcb}{\overline{\Lambda_c}}
\newcommand{\Dstarm}{\mbox{D}^{\ast -}}
\newcommand{\Dsstarp}{\mbox{D}_s^{\ast +}}
\newcommand{\Dsstar}{\mbox{D}^{\ast \ast}}
\newcommand{\Km}{\mbox{K}^-}
\newcommand{\Pb}{P_{b-baryon}}
\newcommand{\KKpi}{\rm{ K K \pi }}
\newcommand{\GeV}{\rm{GeV}}
\newcommand{\GeVc}{\rm{GeV/c}}
\newcommand{\GeVcd}{\rm{GeV/c^2}}
\newcommand{\MeV}{\rm{MeV}}
\newcommand{\MeVc}{\rm{MeV/c}}
\newcommand{\MeVcd}{\rm{MeV/c^2}}
\newcommand{\nb}{\rm{nb}}
\newcommand{\Zzero}{{\rm Z}^0}
\newcommand{\MZ}{\rm{M_Z}}
\newcommand{\MW}{\rm{M_W}}
\newcommand{\GF}{\rm{G_F}}
\newcommand{\Gm}{\rm{G_{\mu}}}
\newcommand{\MH}{\rm{M_H}}
\newcommand{\MT}{\rm{m_{top}}}
\newcommand{\GZ}{\Gamma_{\rm Z}}
\newcommand{\Afb}{\rm{A_{FB}}}
\newcommand{\Afbs}{\rm{A_{FB}^{s}}}
\newcommand{\sigmaf}{\sigma_{\rm{F}}}
\newcommand{\sigmab}{\sigma_{\rm{B}}}
\newcommand{\NF}{\rm{N_{F}}}
\newcommand{\NB}{\rm{N_{B}}}
\newcommand{\Nnu}{\rm{N_{\nu}}}
\newcommand{\RZ}{\rm{R_Z}}
\newcommand{\rhob}{\rho_{eff}}
\newcommand{\Gammanz}{\rm{\Gamma_{Z}^{new}}}
\newcommand{\Gammani}{\rm{\Gamma_{inv}^{new}}}
\newcommand{\Gammasz}{\rm{\Gamma_{Z}^{SM}}}
\newcommand{\Gammasi}{\rm{\Gamma_{inv}^{SM}}}
\newcommand{\Gammaxz}{\rm{\Gamma_{Z}^{exp}}}
\newcommand{\Gammaxi}{\rm{\Gamma_{inv}^{exp}}}
\newcommand{\rhoZ}{\rho_{\rm Z}}
\newcommand{\thw}{\theta_{\rm W}}
\newcommand{\swsq}{\sin^2\!\thw}
\newcommand{\swsqmsb}{\sin^2\!\theta_{\rm W}^{\overline{\rm MS}}}
\newcommand{\swsqbar}{\sin^2\!\overline{\theta}_{\rm W}}
\newcommand{\cwsqbar}{\cos^2\!\overline{\theta}_{\rm W}}
\newcommand{\swsqb}{\sin^2\!\theta^{eff}_{\rm W}}
\newcommand{\ee}{{e^+e^-}}
\newcommand{\eeX}{{e^+e^-X}}
\newcommand{\gaga}{{\gamma\gamma}}
\newcommand{\mumu}{\ifmmode {\mu^+\mu^-} \else ${\mu^+\mu^-} $ \fi}
\newcommand{\eeg}{{e^+e^-\gamma}}
\newcommand{\mumug}{{\mu^+\mu^-\gamma}}
\newcommand{\tautau}{{\tau^+\tau^-}}
\newcommand{\qqb}{{q\bar{q}}}
\newcommand{\eegg}{e^+e^-\rightarrow \gamma\gamma}
\newcommand{\eeggg}{e^+e^-\rightarrow \gamma\gamma\gamma}
\newcommand{\eeee}{e^+e^-\rightarrow e^+e^-}
\newcommand{\eeeeee}{e^+e^-\rightarrow e^+e^-e^+e^-}
\newcommand{\eeeeg}{e^+e^-\rightarrow e^+e^-(\gamma)}
\newcommand{\eeeegg}{e^+e^-\rightarrow e^+e^-\gamma\gamma}
\newcommand{\eeeg}{e^+e^-\rightarrow (e^+)e^-\gamma}
\newcommand{\eemumu}{e^+e^-\rightarrow \mu^+\mu^-}
\newcommand{\eetautau}{e^+e^-\rightarrow \tau^+\tau^-}
\newcommand{\eehad}{e^+e^-\rightarrow {\rm hadrons}}
\newcommand{\eettg}{e^+e^-\rightarrow \tau^+\tau^-\gamma}
\newcommand{\eell}{e^+e^-\rightarrow l^+l^-}
\newcommand{\Ztopig}{{\rm Z}^0\rightarrow \pi^0\gamma}
\newcommand{\Ztogg}{{\rm Z}^0\rightarrow \gamma\gamma}
\newcommand{\Ztoee}{{\rm Z}^0\rightarrow e^+e^-}
\newcommand{\Ztoggg}{{\rm Z}^0\rightarrow \gamma\gamma\gamma}
\newcommand{\Ztomumu}{{\rm Z}^0\rightarrow \mu^+\mu^-}
\newcommand{\Ztotautau}{{\rm Z}^0\rightarrow \tau^+\tau^-}
\newcommand{\Ztoll}{{\rm Z}^0\rightarrow l^+l^-}
\newcommand{\Ztocc}{{\rm Z^0\rightarrow c \bar c}}
\newcommand{\Lamp}{\Lambda_{+}}
\newcommand{\Lamm}{\Lambda_{-}}
\newcommand{\Pt}{\rm P_{t}}
\newcommand{\Gee}{\Gamma_{ee}}
\newcommand{\Gpig}{\Gamma_{\pi^0\gamma}}
\newcommand{\Ggg}{\Gamma_{\gamma\gamma}}
\newcommand{\Gggg}{\Gamma_{\gamma\gamma\gamma}}
\newcommand{\Gmumu}{\Gamma_{\mu\mu}}
\newcommand{\Gtautau}{\Gamma_{\tau\tau}}
\newcommand{\Ginv}{\Gamma_{\rm inv}}
\newcommand{\Ghad}{\Gamma_{\rm had}}
\newcommand{\Gnu}{\Gamma_{\nu}}
\newcommand{\GnuSM}{\Gamma_{\nu}^{\rm SM}}
\newcommand{\Gll}{\Gamma_{l^+l^-}}
\newcommand{\Gff}{\Gamma_{f\overline{f}}}
\newcommand{\Gtot}{\Gamma_{\rm tot}}
\newcommand{\Rb}{\mbox{R}_b}
\newcommand{\Rc}{\mbox{R}_c}
\newcommand{\al}{a_l}
\newcommand{\vl}{v_l}
\newcommand{\af}{a_f}
\newcommand{\vf}{v_f}
\newcommand{\ael}{a_e}
\newcommand{\ve}{v_e}
\newcommand{\amu}{a_\mu}
\newcommand{\vmu}{v_\mu}
\newcommand{\atau}{a_\tau}
\newcommand{\vtau}{v_\tau}
\newcommand{\ahatl}{\hat{a}_l}
\newcommand{\vhatl}{\hat{v}_l}
\newcommand{\ahate}{\hat{a}_e}
\newcommand{\vhate}{\hat{v}_e}
\newcommand{\ahatmu}{\hat{a}_\mu}
\newcommand{\vhatmu}{\hat{v}_\mu}
\newcommand{\ahattau}{\hat{a}_\tau}
\newcommand{\vhattau}{\hat{v}_\tau}
\newcommand{\vtildel}{\tilde{\rm v}_l}
\newcommand{\avsq}{\ahatl^2\vhatl^2}
\newcommand{\Ahatl}{\hat{A}_l}
\newcommand{\Vhatl}{\hat{V}_l}
\newcommand{\Afer}{A_f}
\newcommand{\Ael}{A_e}
\newcommand{\Aferb}{\bar{A_f}}
\newcommand{\Aelb}{\bar{A_e}}
\newcommand{\AVsq}{\Ahatl^2\Vhatl^2}
\newcommand{\Iwk}{I_{3l}}
\newcommand{\Qch}{|Q_{l}|}
\newcommand{\roots}{\sqrt{s}}
\newcommand{\pT}{p_{\rm T}}
\newcommand{\mt}{m_t}
\newcommand{\Rechi}{{\rm Re} \left\{ \chi (s) \right\}}
\newcommand{\up}{^}
\newcommand{\abscosthe}{|cos\theta|}
\newcommand{\dsum}{\Sigma |d_\circ|}
\newcommand{\zsum}{\Sigma z_\circ}
\newcommand{\sint}{\mbox{$\sin\theta$}}
\newcommand{\cost}{\mbox{$\cos\theta$}}
\newcommand{\mcost}{|\cos\theta|}
\newcommand{\dgsl}{{\rm d}\Gamma (\Bdb \rightarrow \Dstarp \ell^-
\overline{\nu}_{\ell})}
\newcommand{\epair}{\mbox{$e^{+}e^{-}$}}
\newcommand{\mupair}{\mbox{$\mu^{+}\mu^{-}$}}
\newcommand{\taupair}{\mbox{$\tau^{+}\tau^{-}$}}
\newcommand{\gamgam}{\mbox{$e^{+}e^{-}\rightarrow e^{+}e^{-}\mu^{+}\mu^{-}$}}
\newcommand{\fullskip}{\vskip 16cm}
\newcommand{\halfskip}{\vskip  8cm}
\newcommand{\quarskip}{\vskip  6cm}
\newcommand{\abitskip}{\vskip 0.5cm}
\newcommand{\ba}{\begin{array}}
\newcommand{\ea}{\end{array}}
\newcommand{\bc}{\begin{center}}
\newcommand{\ec}{\end{center}}
\newcommand{\be}{\begin{eqnarray}}
\newcommand{\eeq}{\end{eqnarray}}
\newcommand{\bes}{\begin{eqnarray*}}
\newcommand{\ees}{\end{eqnarray*}}
\newcommand{\Kz}{\ifmmode {\rm K^0_s} \else ${\rm K^0_s} $ \fi}
\newcommand{\Zz}{\ifmmode {\rm Z} \else ${\rm Z } $ \fi}
\newcommand{\qqbar}{\ifmmode {\rm q\bar{q}} \else ${\rm q\bar{q}} $ \fi}
\newcommand{\ccbar}{\ifmmode {\rm c\bar{c}} \else ${\rm c\bar{c}} $ \fi}
\newcommand{\bbbar}{\ifmmode {\rm b\bar{b}} \else ${\rm b\bar{b}} $ \fi}
\newcommand{\xxbar}{\ifmmode {\rm x\bar{x}} \else ${\rm x\bar{x}} $ \fi}
\newcommand{\rphi}{\ifmmode {\rm R\phi} \else ${\rm R\phi} $ \fi}
\newcommand{\mysection}[1]{\section{\boldmath #1}}
\newcommand{\mysubsection}[1]{\subsection[#1]{\boldmath #1}}
\newcommand{\mysubsubsection}[1]{\subsubsection[#1]{\boldmath #1}}


\begin{titlepage}

\pagenumbering{arabic}
\vspace*{-1.5cm}
\vspace*{2.cm}
\begin{center}
\Large 
{\bf Extraction of the $x$-dependence of the 
non-perturbative QCD $b$-quark fragmentation distribution component.} 
\vspace*{2.cm}
\\
\normalsize {    {\bf 
E. Ben-Haim$^{1,2}$, Ph. Bambade$^{1}$, P. Roudeau$^{1}$, A. Savoy-Navarro$^{2,3}$\\
and A. Stocchi$^{1}$ }}\\
\vskip 0.5truecm
   {\footnotesize  (1) U. de Paris-Sud, Lab. de l'Acc\'el\'erateur Lin\'eaire (L.A.L.), Orsay and IN2P3-CNRS } \\ 
   {\footnotesize  (2) LPNHE, Universit\'es de Paris 6-7 and IN2P3-CNRS} \\
   {\footnotesize  (3) Also Visitor at Fermilab in the CDF experiment} \\
\end{center}
\vspace{\fill}

\begin{abstract}
\noindent
Using recent measurements of the $b$-quark fragmentation distribution obtained in $\epem \rightarrow b \overline{b}$ events registered at the $\Zz$ pole, the non-perturbative QCD component of the distribution has been extracted
independently of any hadronic physics modelling.
This distribution depends only on the way the perturbative QCD component 
has been defined. When the perturbative QCD component is taken from
a parton shower Monte-Carlo, the non-perturbative QCD component
is rather similar with those obtained from the Lund or Bowler models.
When the perturbative QCD component is the result of an analytic NLL
computation, the non-perturbative QCD component has to be extended
in a non-physical region and thus cannot be described by any 
hadronic modelling. In the two examples used to characterize
these two situations, which are studied at present, it happens
that the extracted non-perturbative QCD distribution has the same shape,
being simply translated to higher-$x$ values in the second approach,
illustrating the ability of the analytic 
perturbative QCD approach to account for
softer gluon radiation than with a parton shower generator.
\end{abstract}

\vspace{\fill}
\end{titlepage}

\setcounter{page}{1}    

\mysection {Introduction}
\label{sec:intro}
Improved determinations of the $b$-quark fragmentation distribution
have been obtained by ALPEH \cite{ref:aleph1}, DELPHI \cite{ref:delphi1}, 
OPAL \cite{ref:opal1} and SLD \cite{ref:sld1} collaborations
which measured the fraction of the beam energy taken by a weakly
decaying $b$-hadron in $\epem \rightarrow b \overline{b}$ events
registered at, or near, the $\Zz$ pole.

This distribution is generally viewed as resulting from three
components: the primary interaction ($\epem$ annihilation into 
a $b \overline{b}$ pair in the present study), a perturbative
QCD description of gluon emission by the quarks and a non-perturbative
QCD component which incorporates all mechanisms at work to bridge 
the gap between the previous phase and the production of weakly
decaying $b$-mesons. The perturbative QCD component can be obtained 
using analytic expressions or Monte Carlo generators. The non-perturbative QCD component is usually parametrized phenomenologically via a model.

To compare with experimental results, one must fold both components to  evaluate the expected $x$-dependence\footnote{In the present analysis, 
$x=\frac{\sqrt{x_{E}^{2}-x_{min}^{2}}}{\sqrt{1-x_{min}^{2}}}$
where $x_{E}=\frac{2E_{B}}{\sqrt{s}}$ is the fraction of the beam energy
 taken by the weakly decaying hadron and $x_{min}=\frac{2m_{B}}{\sqrt{s}}$
 is its minimal value.}:
\begin{equation}
{\cal D}_{predicted}(x) = \int_0^1{ {\cal D}_{pert.}(z) \times  
{\cal D}_{non-pert.}^{model}(\frac{x}{z})~\frac{dz}{z}}
\label{eq:start}
\end{equation}
The final and the perturbative components are defined over the $[0,1]$ interval.
As explained, in the following, the non-perturbative distribution must be evaluated
for $x>1$, if the perturbative component is non-physical. 
The parameters of the model are then fitted by comparing the measured and predicted
$x$-dependence of the $b$-quark fragmentation distribution.
Such comparisons have already been made by the different experiments
using, for the
perturbative component, expectations from generators such as the 
JETSET or HERWIG  parton shower Monte-Carlo.
It has been shown, with present measurement accuracy, that most of existing 
models, for the non-perturbative part,
are unable to give a reasonable fit to the data 
\cite{ref:aleph1,ref:delphi1,ref:opal1,ref:sld1}. Best results have been
obtained with the Lund and Bowler models \cite{ref:lund1,ref:bowler1}.

In the following, a method is presented to extract the 
non-perturbative QCD component of the fragmentation function 
directly from data,
independently of any hadronic model assumption. 
This distribution can then be 
compared with models to learn about the non-perturbative
QCD transformation of $b$-quarks into $b$-hadrons.
It can then be used
in another environment than $\epem$ annihilation, as long as the same
parameters and methods are taken for the evaluation of
the perturbative QCD component. Consistency checks,
on the matching between the measured and predicted $b$-fragmentation
distribution, can be defined which provide information on the determination
of the perturbative QCD component itself.

In Section \ref{sec:extract}, the method used to extract 
the non-perturbative QCD component is presented.

In Section \ref{sec:meast}, the extraction is performed for two 
determinations of the perturbative QCD component using: 
\begin{itemize}
\item the JETSET 7.3 generator \cite{ref:jetset1}, tuned on  DELPHI data \cite{ref:tune1}, running in the parton shower mode, 

\item an analytic computation based on QCD at NLL order \cite{ref:catani1}.

\end{itemize}

In Section \ref{sec:analyse} these results are discussed and insights obtained with the present analysis are explained. A  parametrization for the non-perturbative QCD component is proposed.

\mysection{\mbox{Extracting the $x$-dependence of the non-perturbative} \\ QCD component}
\label{sec:extract}

The method is based on the use of the Mellin transformation which is 
appropriate when dealing with integral equations as given in
(\ref{eq:start}).
The Mellin transformation of the expression for ${\cal D}(x)$
is:

\begin{equation}
\tilde{{\cal D}}(N) = \int_0^{\infty}{dx ~x^{N-1}~{\cal D}(x)}
\label{eq:mellint}
\end{equation}
where $N$ is a complex variable. For integer values
of $ N \ge 2$, the values of $\tilde{{\cal D}}(N)$ correspond to the moments
of the initial $x$ distribution \footnote{By definition  
$\tilde{{\cal D}}(1)~(= 1)$ corresponds to the normalization of 
${\cal D}(x)$.}. For physical processes, $x$ is restricted to be within
the $[0,1]$ interval.
The interest in using Mellin transformed expressions is that
Equation (\ref{eq:start}) becomes a simple product:
\begin{equation}
\tilde{{\cal D}}(N) = \tilde{{\cal D}}_{pert.}(N) \times  
\tilde{{\cal D}}_{non-pert.}(N)
\label{eq:mellintb}
\end{equation}

Having computed, in the $N$-space, distributions of the measured
and perturbative QCD components, the non-perturbative distribution,
$\tilde{{\cal D}}_{non-pert.}(N)$ is obtained from Equation 
(\ref{eq:mellintb}). Applying the inverse Mellin transformation
on this distribution one gets ${\cal D}_{non-pert.}(x)$
without any need for a model input:

\begin{equation}
{\cal D}_{non-pert.}(x) =
\frac{1}{2\pi i}\oint{dN~\frac{\tilde{{\cal D}}_{meas.}(N)}{\tilde{{\cal D}}_{pert.}(N)}~x^{-N}}
\label{eq:inv_mellin}
\end{equation}
in which the integral runs over a contour in the complex $N$-plane. The integration 
contour is taken as two symmetric straight half-lines, 
one in the upper half and the other in the 
lower half of the complex plane. 
The angle of the lines, relative to the real axis, is larger or smaller than   
90 degrees for $x$ values smaller or larger than unity, respectively.
These lines are taken to originate from
$N=(1.01,0)$.
The contour is supposed to be closed by an arc 
situated at infinity in the negative and positive directions of the real axis,
for the two cases respectively  
\footnote {It has been verified that the result is independent of a
definite choice for the contour in terms of the slope
of the lines and of the value of the arc radius. The result is also
independent of the choice for the position of the origin of the lines,
on the real axis, as long as the contour encloses the singularities
of the expression to be integrated and stays away from the Landau
pole present in $\tilde{{\cal D}}_{pert.}(N)$ which is discussed
in the following.}. 

\begin{figure}[htb!]
  \begin{center}
    \mbox{
\epsfig{file=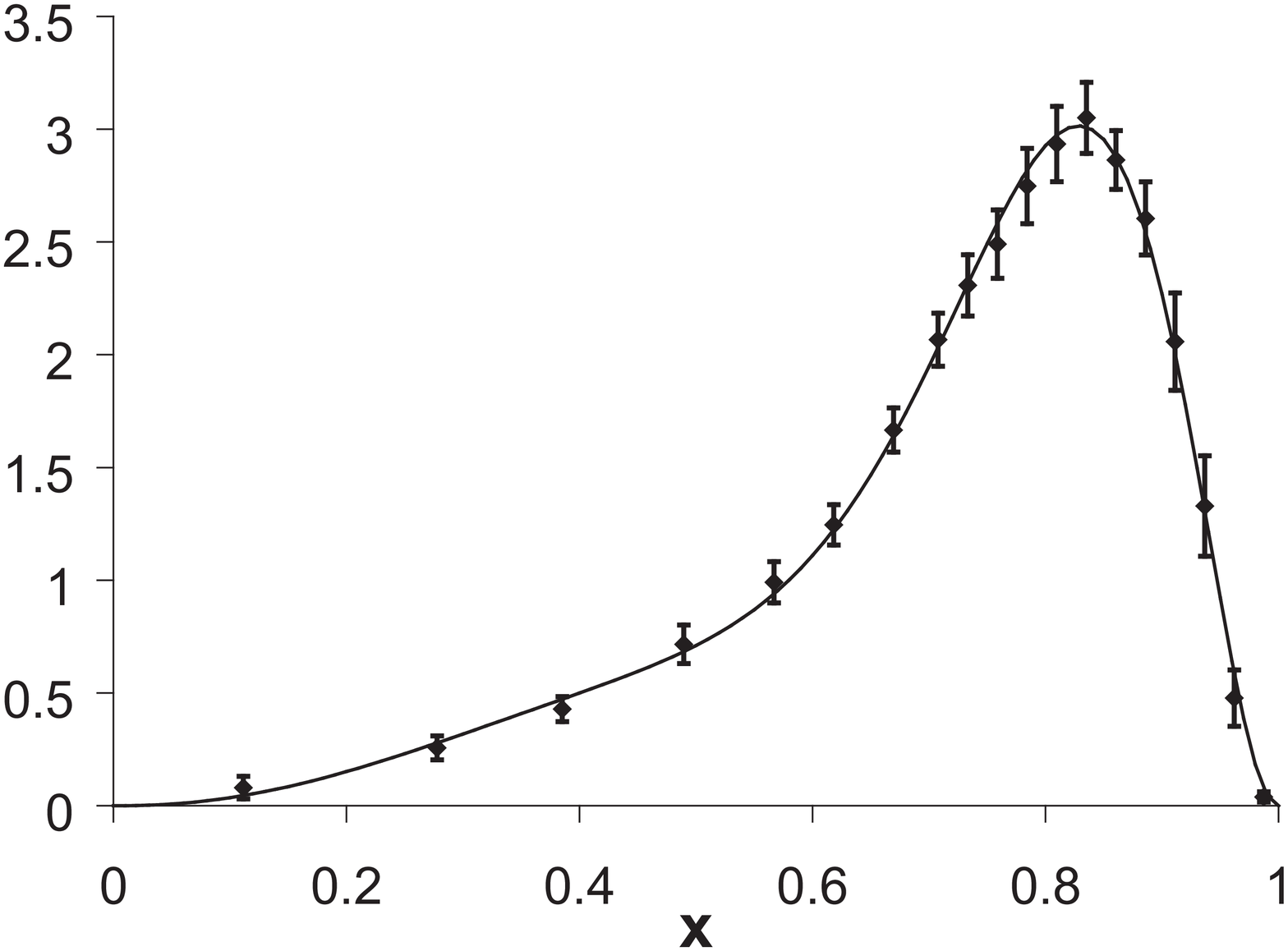,width=7cm,height=7cm},
\epsfig{file=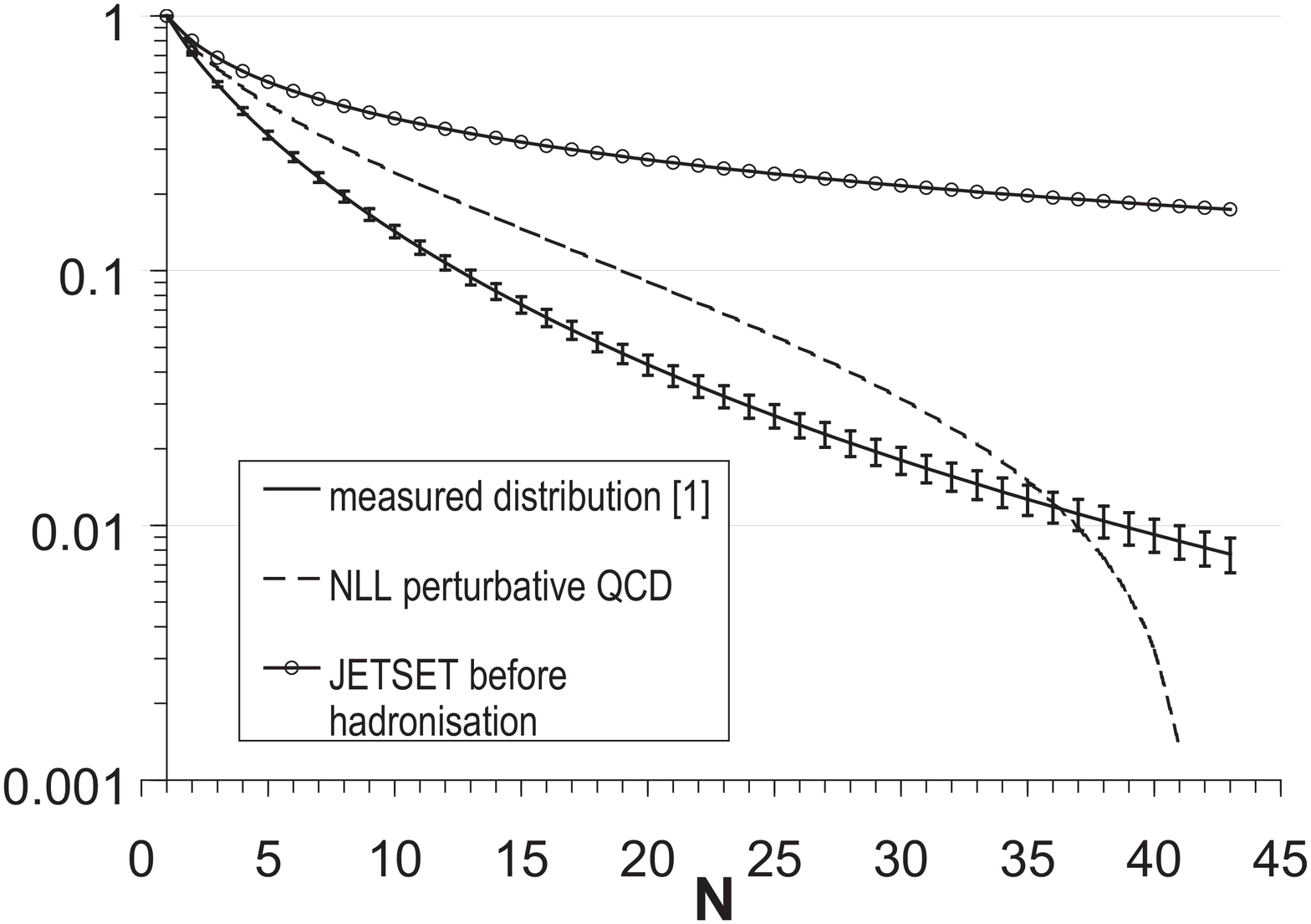,width=7cm,height=7cm}}
  \end{center}
  \caption[]{\it {
Left: Comparison between the measured (points with error bars)
$b$-fragmentation distribution 
and the fitted parametrization using Equation (\ref{eq:eqdata}).
Right: Moments of the measured (full line)
$x$ distribution, of the perturbative QCD component \cite{ref:catani1}(dashed line)
and of the generated distribution obtained in JETSET before hadronization
(full line with circles). Data from \cite{ref:aleph1} have been used. 
}
   \label{fig:mom}}
\end{figure}
In practice, the Mellin transformed distribution of present measurements,
$\tilde{{\cal D}}_{meas.}(N)$, has been 
obtained
after having adjusted an analytic expression to the measured distribution in $x$,
and by applying the Mellin transformation on this fitted function. 
The following expression, which depends
on five parameters and gives a good description of the measurements
(see Figure \ref{fig:mom}-left),
has been used.
\begin{equation}
D(x)=p_0 \times \left [ p_1 x^{p_2}(1-x)^{p3}+(1-p_1)x^{p_4}(1-x)^{p5}\right ]
\label{eq:eqdata}
\end{equation}
where $p_0$ is a normalisation coefficient.
Values of the parameters have been obtained by comparing, in each bin,
the measured bin content with the integral of $D(x)$ over the bin.

Measurements of the $b$-fragmentation distribution, in \cite{ref:aleph1}, 
have been published in a binned form, after unfolding of the experimental 
energy resolution. Values in the bins are correlated and,
as the bin width is smaller than 
the resolution, the error matrix is singular.  Only positive
eigenvalues of this matrix have been used in the present
analysis, when fitting parameters.


The distribution of moments
obtained with data from \cite{ref:aleph1}, and computed using
the fitted distribution corresponding to Equation (\ref{eq:eqdata}), is given in 
Figure \ref{fig:mom}-Right. 
The corresponding analytic expression is:
\begin{equation}
\tilde{{\cal D}}(N)=p_0 \left [
p_1 \frac{\Gamma (p_2+N)}{\Gamma (p_2+p_3+N+1)}
+(1-p_1)\frac{\Gamma (p_4+N)}{\Gamma (p_4+p_5+N+1)}
\right ]
\label{eq:momdata}
\end{equation}
Quoted uncertainties, in Figure \ref{fig:mom}-Right, correspond to 
actual measurements
and are highly correlated. 
They have been obtained by propagating uncertainties corresponding to 
the covariance matrix of the $p_{1,.,5}$ fitted parameters.
When computing the moments,
very similar results are obtained, for $N<10$, using directly the 
measured $x$-binned distribution. For higher $N$ values 
effects induced by the variation
of the distribution within a bin, as expressed by Equation (\ref{eq:eqdata}),
have to be included.

The Mellin transformed distribution of the JETSET perturbative QCD component has 
been obtained in a similar way, whereas the NLL QCD perturbative component is 
computed directly as a function of $N$ in \cite{ref:catani1}.
At large values of $N$, this last distribution is equal to zero for 
$N=N_0\simeq 41.7$ and has a Landau pole situated at $N_L\simeq 44.$
\footnote{Values for $N_0$ and $N_L$ depend on the exact values
assumed for the other parameters entering into the 
computation; see Section \ref{sec:subsecth} where values
of these parameters have been listed.}

\mysection {\mbox{$x$-dependence measurement of the non-perturbative} QCD
component}
\label{sec:meast}

The $x$ distribution of the  non-perturbative QCD component
extracted in this way depends on the measurements and also
on the procedures adopted to compute the perturbative QCD component.
In the following two approaches have been considered. The first one is generally
adopted by experimentalists whereas the second is more frequent
for theorists.

\begin{figure}[htb!]
  \begin{center}
    \mbox{\epsfig{file=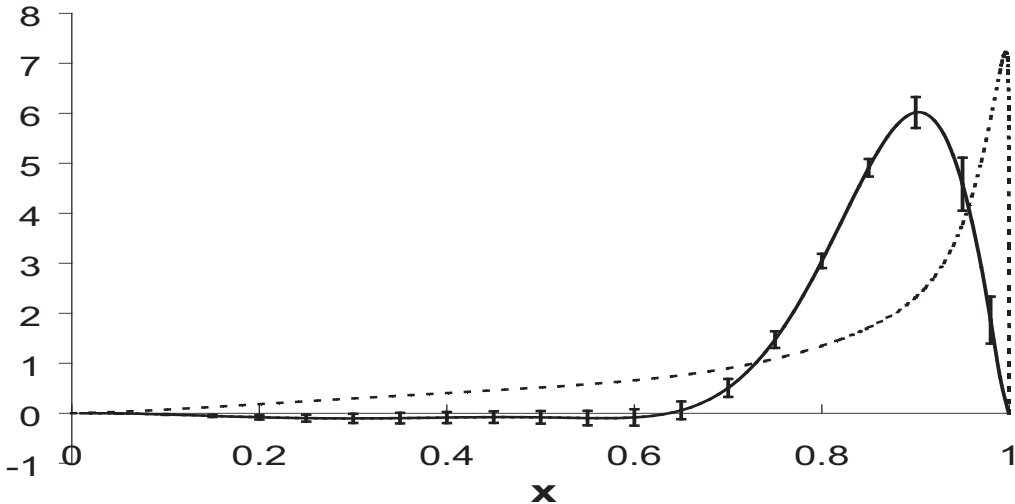,
width=14cm,height=7cm}}
  \end{center}
  \caption[]{\it {$x$-dependence of the perturbative (dotted line) and
non-perturbative (full line) QCD components of the measured \cite{ref:aleph1}
$b$-fragmentation
distribution. 
These curves are obtained by interpolating corresponding values determined 
at a large number of points in the x-variable. 
Quoted error bars correspond to measurement uncertainties and are correlated for different $x$-values. The perturbative QCD component is extracted from the JETSET 7.3 Monte Carlo generator. 
The dotted curve has to be complemented by a $\delta$-function containing 4$\%$ 
of the events, located at $x=1$.} 
   \label{fig:jetset}}
\end{figure}

\mysubsection{The perturbative QCD component is provided by a
generator}
\label{sec:gen}
The JETSET 7.3 Monte Carlo generator, with values of the parameters
tuned on DELPHI data registered at the $\Zz$ pole has been used.
Events have been produced using the parton shower option of the generator
and the $b$-quark energy is extracted, after radiation of gluons, just
before calling the routines to create a $b$-hadron that takes a fraction $z$
\footnote{$z=\frac{E^B+p_L^B}{E^b+p_L^b}$ is the boost-invariant
fraction of the $b$-jet energy taken by the weakly decaying B meson. This variable is defined for a string stretched between the $b$-quark and a gluon, an anti-quark or a diquark. In the present analysis no distinction has been introduced between $z$ and $x$ as no string model has been considered.}
of the available string energy. 
This distribution is displayed in Figure \ref{fig:jetset}. It has 
to be complemented by a $\delta$-function
at $x=1$ which contains $\sim 4\%$ of all events. In this peak, $b$-hadrons
carry all the energy of the $b$-quark as no gluon has
been radiated.

Applying the method explained in Section \ref{sec:extract}, the corresponding 
non-perturbative QCD component has been extracted, and  is displayed also
in Figure \ref{fig:jetset}. 
Above $x=1$, it is compatible with zero, as expected.
The quoted error bar, for a given value of $x$, has been obtained by evaluating
the values of ${\cal D}_{non-pert.}(x)$ for different shapes of the 
$b$-quark fragmentation distribution which are obtained by varying
parameters $p_{1,.,5}$ according to their measured error matrix.


\mysubsection{The perturbative QCD component is obtained by an analytic
computation based on QCD}
\label{sec:subsecth}

\begin{figure}[htb!]
  \begin{center}
    \mbox{\epsfig{file=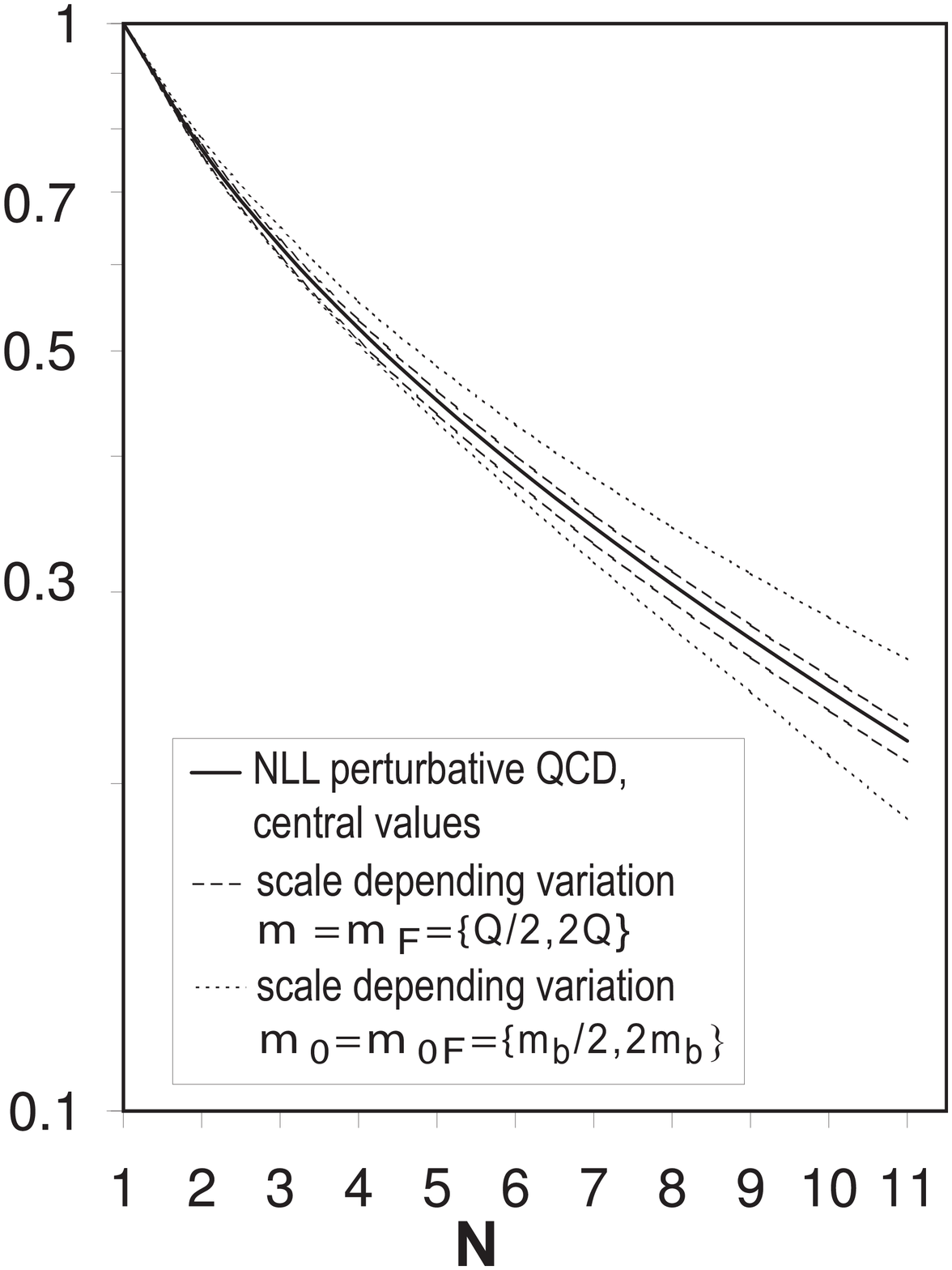, 
width=7cm,height=7cm}
          \epsfig{file=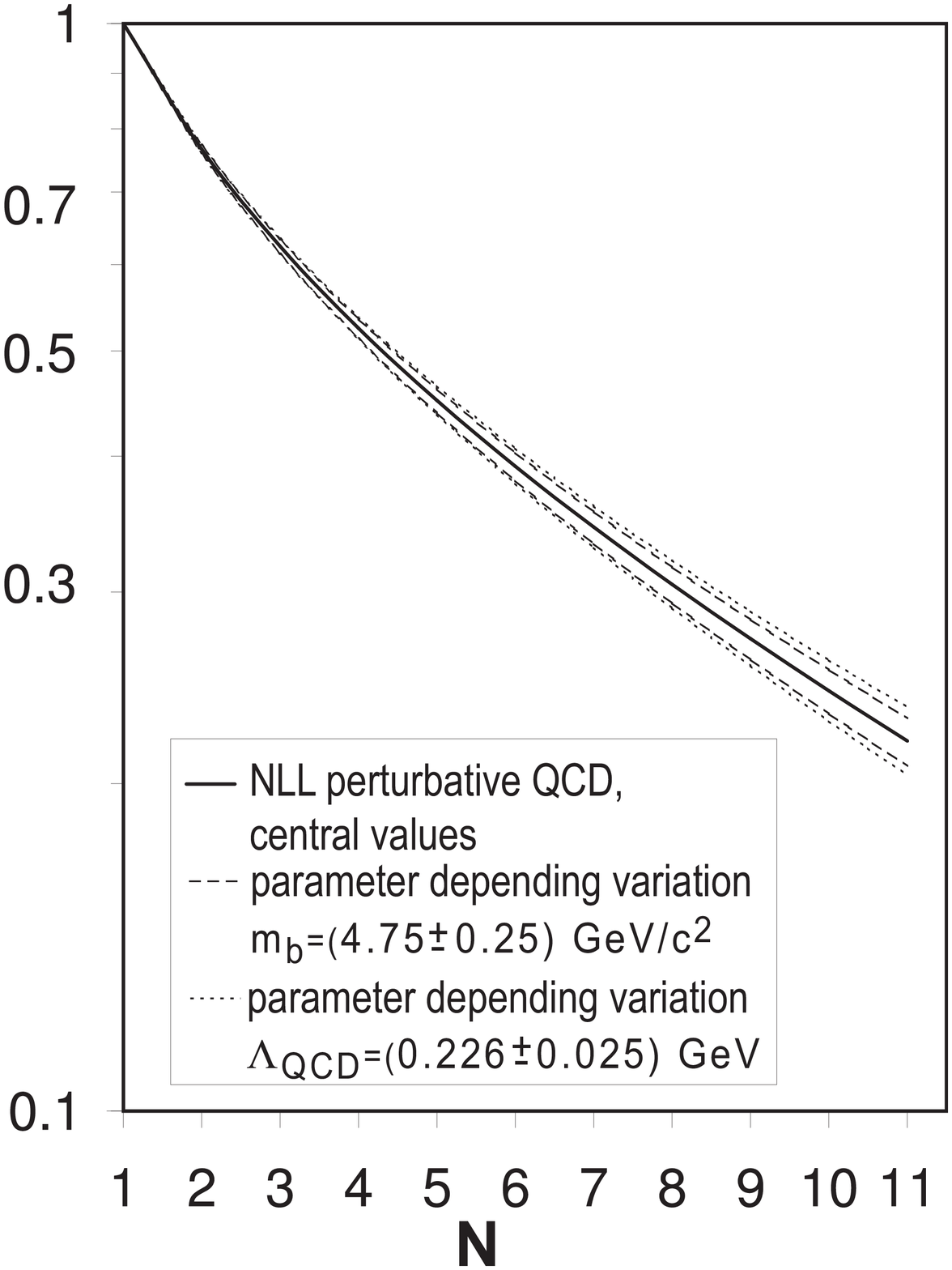, 
width=7cm,height=7cm}
}

  \end{center}
  \caption[]{\it Variations of the calculated perturbative QCD component \cite{ref:catani1}, depending on the renormalization scales ($\mu$, $\mu_0$), the factorisation scales ($\mu_F$, $\mu_{0F}$), the $b$-quark mass ($m_b$) and $\Lambda^{(5)}_{QCD}$. The full lines are corresponding to the central values ($\mu=\mu_{F}=Q=91.2 ~\GeV$, $\mu_{0}=\mu_{0F}=m_{b}$, $m_{b}=4.75 ~\GeV/c^2$ and $\Lambda^{(5)}_{QCD}=0.226~ \GeV$). 
}
   \label{fig:mom_scale}
\end{figure}

The perturbative QCD fragmentation function is evaluated according to 
the approach 
presented in  \cite{ref:catani1}. 
This next to leading log (NLL) accuracy calculation for the inclusive $b$-quark 
production cross section  in $\epem$ annihilation, generalises previous 
calculations 
by resumming the contribution from soft gluon radiation
(which plays an important role at large $x$) to all
perturbative orders and to NLL accuracy.
These computations are done directly in the $N$-space.
        Soft gluon radiation contributes to the logarithm
of the fragmentation function large logarithmic terms 
of the type $(\log{N})^p$, with $p \le n+1$. These terms
appear at all  perturbative 
orders $n$ in $\alpha_s$. In the calculation at  NLL accuracy \cite{ref:catani1}, 
the two largest terms, 
corresponding to $p=n+1~{\rm and}~n$, 
have been resummed at all perturbative orders. 
The calculation is expected to be reliable when $N$ is not too 
large (typically less than 
20). 
To obtain 
distributions for the variable $x$ from results in moment space,
one should apply the inverse Mellin transformation, that consists 
in integrating over a contour in $N$ (Section \ref{sec:extract}). 
When $x$ gets closer to 1, large values of 
$N$ contribute and thus the perturbative fragmentation distribution
is not reliable in these regions.
This behaviour affects also values of the distribution at lower $x$ 
as moments of this distribution are fixed.
In addition to the break-down of the theory for large values of $N$, uncertainties attached to the determination of the theoretical perturbative QCD component are related to the definition of the scales entering into the computation. This component also depends on two parameters: the $b$-quark pole mass ($m_b$) and $\Lambda^{(5)}_{QCD}$, that have been taken as $m_{b}^{pole}=(4.75\pm0.25)~ \GeV/c^2$ and $\Lambda^{(5)}_{QCD}=(0.226\pm0.025)~ \GeV$. Scale and parameter depending variations of the moments of the perturbative QCD component are given in Figure \ref{fig:mom_scale}. These variations are fully correlated versus $N$.

\begin{figure}[htb!]
  \begin{center}
    \mbox{\epsfig{file=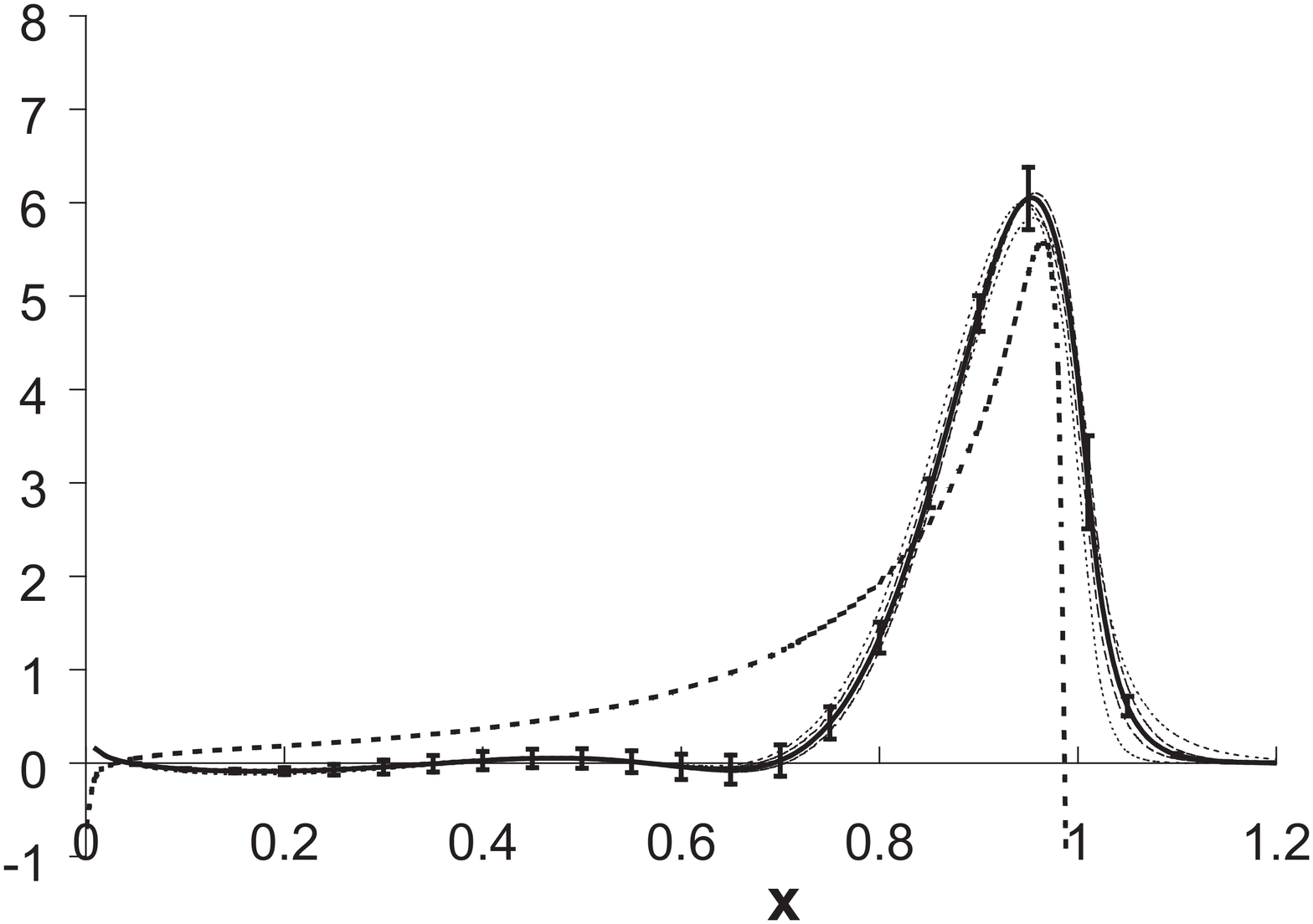,
width=14cm,height=7cm}}
  \end{center}
  \caption[]{\it {$x$-dependence of the perturbative (dotted line) and 
non-perturbative (full line) QCD components of the measured $b$-fragmentation 
distribution. 
These curves are obtained by interpolating corresponding values determined 
at a large number of points in the x-variable.
The perturbative QCD component is given by the analytic computation of \cite{ref:catani1}. The thin lines on both sides of the non-perturbative distribution are corresponding to $\mu_0=\mu_{0F}=\{m_b/2,2m_b\}$ (dotted lines) and $\Lambda^{(5)}_{QCD}=(0.226\pm0.025) ~\GeV$ (dashed lines). Variations induced by 
the other parameters, $\mu=\mu_F=\{Q/2,2Q\}$ and $m_b=(4.75\pm0.25) ~\GeV/c^2$ are smaller. In addition, quoted error bars correspond to measurement uncertainties and are correlated for different $x$-values. 
The perturbative QCD dotted curve has to be complemented by a $\delta$-function
 containing 5$\%$ of the events, located at $x=1$.
}

   \label{fig:npx}}
\end{figure}

 The extracted non-perturbative component 
is given in Figure \ref{fig:npx}. Its shape
depends on the same quantities as those used to evaluate the perturbative distribution, and thus similar variations appear, as drawn also in the Figure. 

It has to be noted that the data description in terms of a product of 
two QCD components, perturbative and non-perturbative, is not directly affected 
by uncertainties attached to the determination of the perturbative component. 
This is because the non-perturbative component, as determined in the present 
approach, compensates for a given choice of method or of parameter values. 


To obtain the complete expected $x$-distribution of $b$-hadrons from the theoretical calculation, one has to be able to evaluate the integral given in Equation (\ref{eq:start}). When $x$ becomes close to 1 there are numerical problems, and consequently, the high-$x$ ($x>0.96$) behaviour of the perturbative QCD component has been studied.
As mentioned before, this region corresponds to high-N values 
where the perturbative approach fails.
As a result, the high-$x$ behaviour of the distribution is non-physical; it oscillates.
To have a numerical control of the distribution 
in this region, it has been decided to take into account
 $x$ values which are below  
a given maximum value, $x_{max}$, above which the
distribution is assumed to be equal to zero.
Moments of this
truncated distribution show a small discrepancy when compared with moments
of the full distribution. 
This difference has an almost linear dependence with $N$. 
To correct for this effect, $x_{max}$
is chosen such that the difference between moments is a constant value
(the slope in $N$ being close to zero at this point). This difference
can then be corrected by adding simply
a $\delta$-function
at $x=1$, so that the total distribution is normalized to 1.
 A typical value for $x_{max}$ is 0.997
and the $\delta$ component corresponds to 5$\%$ of the distribution. We stress that 
the truncation at $x_{max}$ and the added $\delta$-function do not contribute in 
the determination of the non-perturbative component using Equation 
(\ref{eq:inv_mellin}). 
But this procedure is necessary 
for checking that the extracted non-perturbative component in the $x$-space, when 
convoluted with the perturbative distribution, effectively reproduces the 
measurements and also for testing hadronic models given in the $x$-space.

Conversely to the perturbative QCD component which was, by definition
in \cite{ref:catani1}, defined within the $[0,1]$ interval, the
non-perturbative component has to be extended in the region $x>1$.
This ``non-physical'' behaviour
comes from the zero of $\tilde{{\cal D}}_{pert.}(N)$ for $N=N_0$
which gives a pole in the expression to be integrated in 
Equation (\ref{eq:inv_mellin}).
Using properties of integrals in the complex plane, it can be shown
that, for $x>1$, the non-perturbative QCD distribution can be well
approximated by $x^{-N_0}$.

Errors bars, given in Figure \ref{fig:npx}, have been 
obtained using the same procedure as explained in Section \ref{sec:subsecth}.

\mysection{Results interpretation}
\label{sec:analyse}


The $x$-dependence of the non-perturbative QCD component, obtained in this way,
does not depend on any non-perturbative QCD model assumption but its shape is 
tightly 
related to the procedures used to evaluate the perturbative component, and thus, 
the two distributions have to be used jointly. 

\begin{figure}[htb!]
  \begin{center}
    \mbox{\epsfig{file=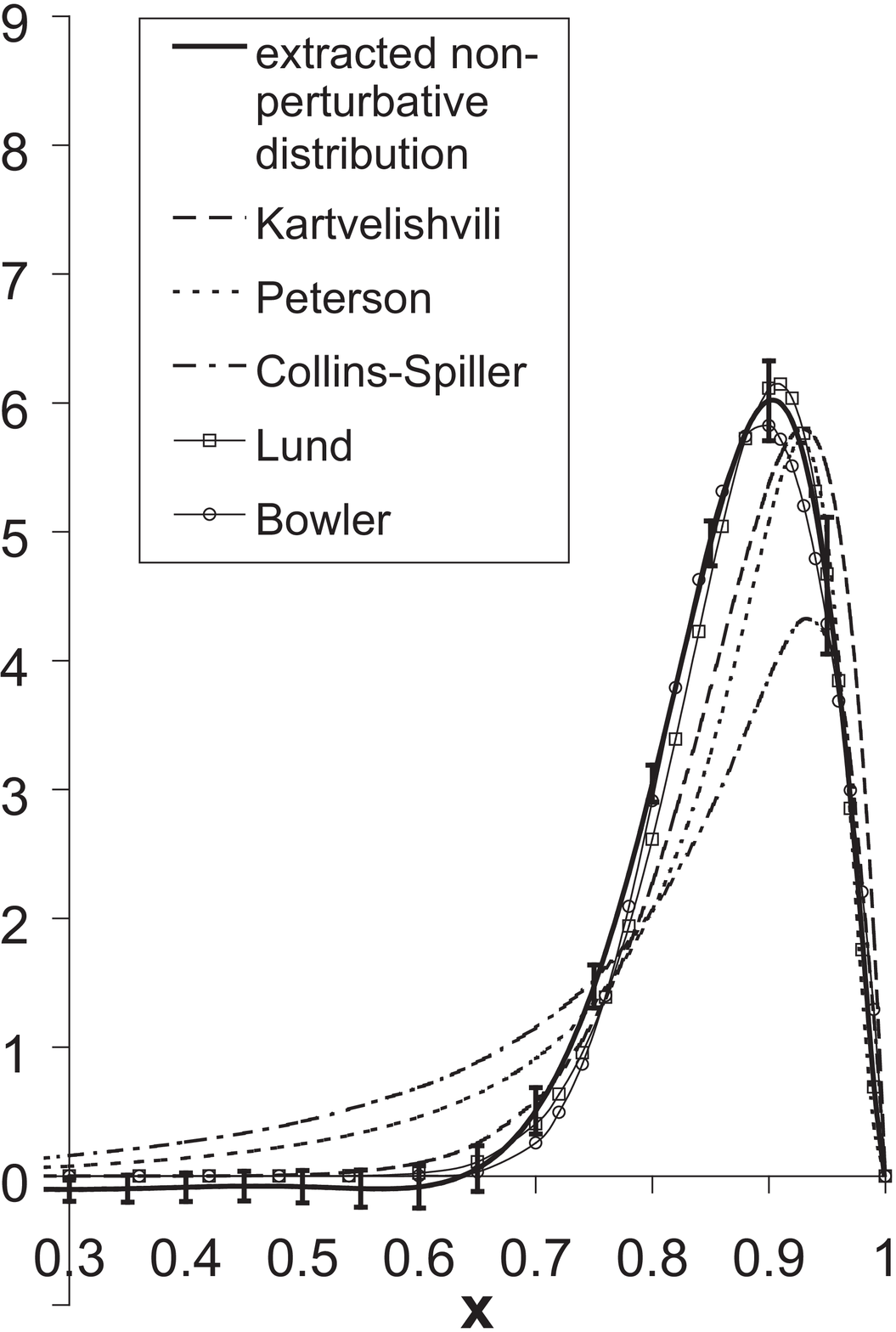, 
width=7cm,height=7cm}
          \epsfig{file=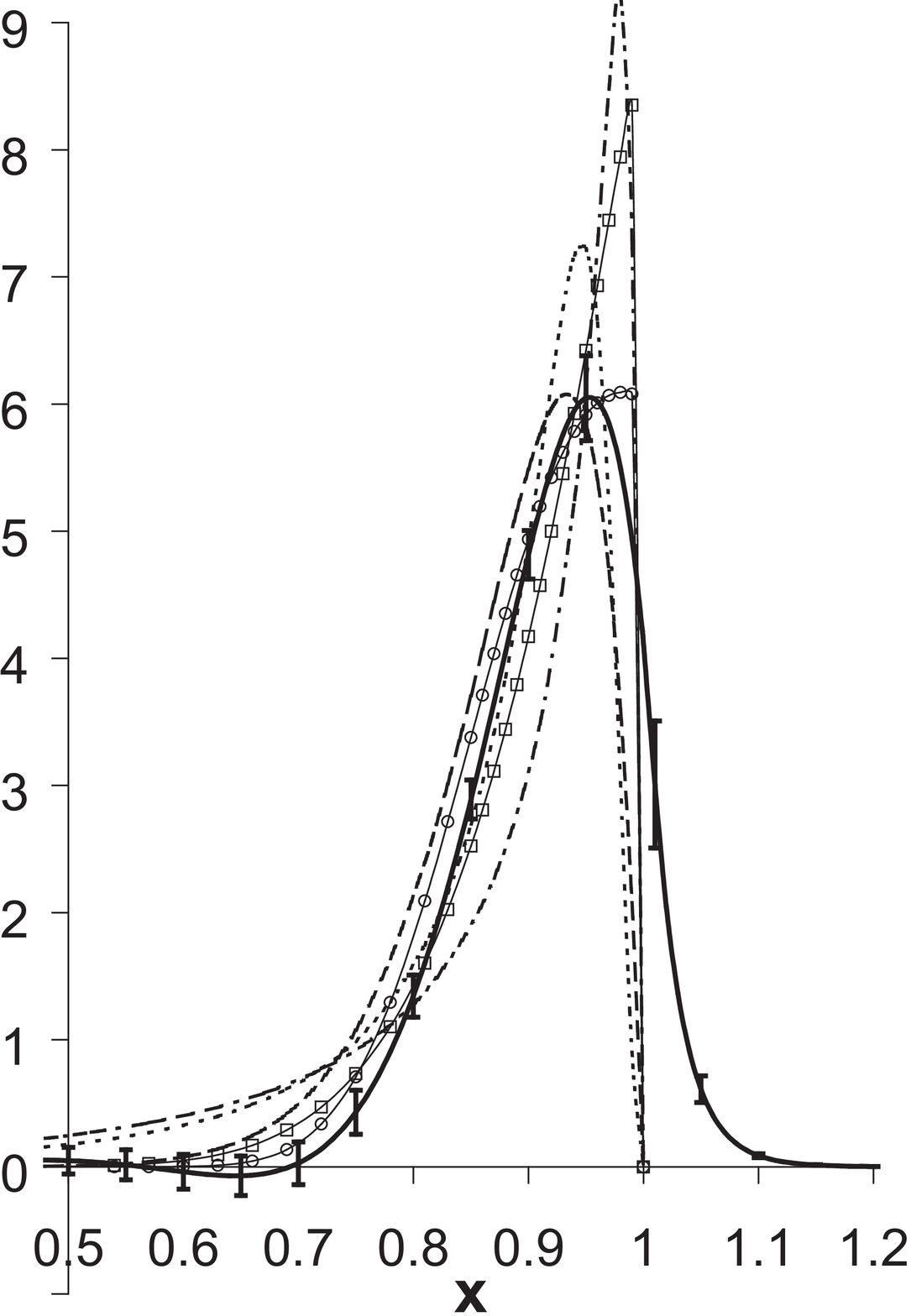,
width=7cm,height=7cm}}

  \end{center}
  \caption[]{\it {Comparison between the directly extracted non-perturbative 
component (thick full line) and the model fits 
on data taken from \cite{ref:aleph1}. 
Left: the perturbative QCD component is taken from JETSET. 
Right: the theoretical perturbative QCD component\cite{ref:catani1} is used.}
   \label{fig:model}}
\end{figure}

\mysubsection{Comparison with models}
Non-perturbative components of the $b$-quark fragmentation
distribution, taken from models, have been folded with a perturbative QCD component
obtained from a Monte Carlo generator and 
compared with measurements by the different collaborations \cite{ref:aleph1, 
ref:delphi1, ref:opal1, ref:sld1}.
Apart for the Lund and Bowler models, 
none of the other models provided a good fit to the data. In Figure 
\ref{fig:model} the directly extracted non-perturbative components are compared
with distributions taken from models \cite{ref:kart1, ref:peterson1, ref:cs1, 
ref:lund1, ref:bowler1} whose parameters have been fitted on data from 
\cite{ref:aleph1}. 
Results have been obtained by comparing, in each bin, the measured bin content
with the integral, over the bin, of the folded expression for 
${\cal D}_{predicted}(x)$. They 
are summarized in Table \ref{tab:fit_res}.
It must be noted that parameters given in this Table, 
when the perturbative QCD component is taken from JETSET, may differ from those
quoted in original publications as an analytic computation is done
in the present study, using Equation (\ref{eq:start}), whereas a string model
is used in the former; other sources of difference can originate from
the exact values of the parameters used to run JETSET and from the definition
of $x$ which varies between 0 and 1 in our case.
Numerically, if one compares present results
with those obtained by OPAL \cite{ref:opal1}, in their own analysis,
for the values of the parameters of the Lund and Bowler models
\footnote{This numerical comparison has been restricted to these two models
as other models do not give a good description of the measurements. 
In addition, only OPAL has fitted the parameters of the two models 
assuming that the transverse $b$-mass was a constant, as done also in our analysis.},
differences
are well within uncertainties.

\begin{table}[htb]
\begin{center}
\tiny{
  \begin{tabular}{|l|c|c|c|c|}
    \hline
~ \hfill Model \hfill ~ & \multicolumn{2}{|c|}{JETSET} & \multicolumn{2}{|c|}{NLL Pert. QCD}\\
    \hline
& param. & $\chi^2/NDF$ & param. & $\chi^2/NDF$ \\ 
    \hline
Kartvelishvili \cite{ref:kart1} &$\epsilon_b=13.2^{+0.8}_{-0.7}$ & $27/4$    & $\epsilon_b=14.0\pm0.7$  & $22/4$ \\
 $x^{\epsilon_b}(1-x)$         &  & & & \\
\hline
Peterson \cite{ref:peterson1}   &$\epsilon_b=(5.6^{+0.7}_{-0.6}) \times 10^{-3}$  & $62/4$   & $\epsilon_b=(3.1 \pm 0.5)\times 10^{-3}$ & $29/4$ \\
 $\frac{1}{x}\left ( 1 - \frac{1}{x}-\frac{\epsilon_b}{1-x} \right )^{-2}$         &  & & & \\
\hline
C.S \cite{ref:cs1}              &$\epsilon_b=(5.4^{+0.9}_{-0.8}) \times 10^{-3}$  & $144/4$   & $\epsilon_b=(4.4^{+1.5}_{-1.1})\times 10^{-4}$   & $18/4$ \\
 $\left (\frac{1-x}{x}+\frac{\epsilon_b(2-x)}{1-x} \right )\left ( 1 +x^2 \right ) \left ( 1 - \frac{1}{x}-\frac{\epsilon_b}{1-x} \right )^{-2}$         &  & & & \\
\hline
Lund \cite{ref:lund1}           &$a=1.55\pm 0.15$                 & $2.6/3$     & $a=0.04 \pm 0.09$             & $0.5/3$ \\
$\frac{1}{x}(1-x)^a \exp{\left ( -\frac{bm_{b\bot}^2}{x}\right )}$ &$bm_{b\bot}^2=14.7^{+1.4}_{-1.2}$       &            & $bm_{b\bot}^2=8.8 ^{+0.9}_{-0.8} $ &         \\
\hline
Bowler \cite{ref:bowler1}       &$a=0.77^{+0.15}_{-0.12}$                 & $2.5/3$     & $a=0.00 \pm 0.03$     & $4.4/3$ \\
$\frac{1}{x^{1+bm_{b\bot}^2}}(1-x)^a \exp{\left ( -\frac{bm_{b\bot}^2}{x}\right )}$ &$bm_{b\bot}^2=63.7^{+6.8}_{-6.0}$       &            & $bm_{b\bot}^2=53.\pm 6.$&         \\
    \hline
  \end{tabular}
}
  \caption[]{\it {Values of the parameters and of the $\chi^2/NDF$ obtained when fitting 
results from Equation (\ref{eq:start}), obtained for different models
of the non-perturbative QCD component, to the measured
$b$-fragmentation distribution. 
The two 
situations corresponding, respectively, to the perturbative QCD component taken
from JETSET or from \cite{ref:catani1} have been distinguished. The Lund and 
Bowler models have been simplified by assuming that the transverse mass of 
the $b$-quark, $m_{b\bot}$, is  a constant.}
  \label{tab:fit_res}}
\end{center}
\end{table}



When the perturbative QCD component
is taken from the analytic NLL calculation the Lund and Bowler models
give also the best description of the measurements.
Values for the parameter $a$ in these models are 
even compatible with zero corresponding to a behaviour
in $1/z^{\alpha} ~ \exp{(-A/z)}$, to accomodate the non-zero value of the non-perturbative
QCD component at $x=1$.

\begin{figure}[htb!]
  \begin{center}
    \mbox{\epsfig{file=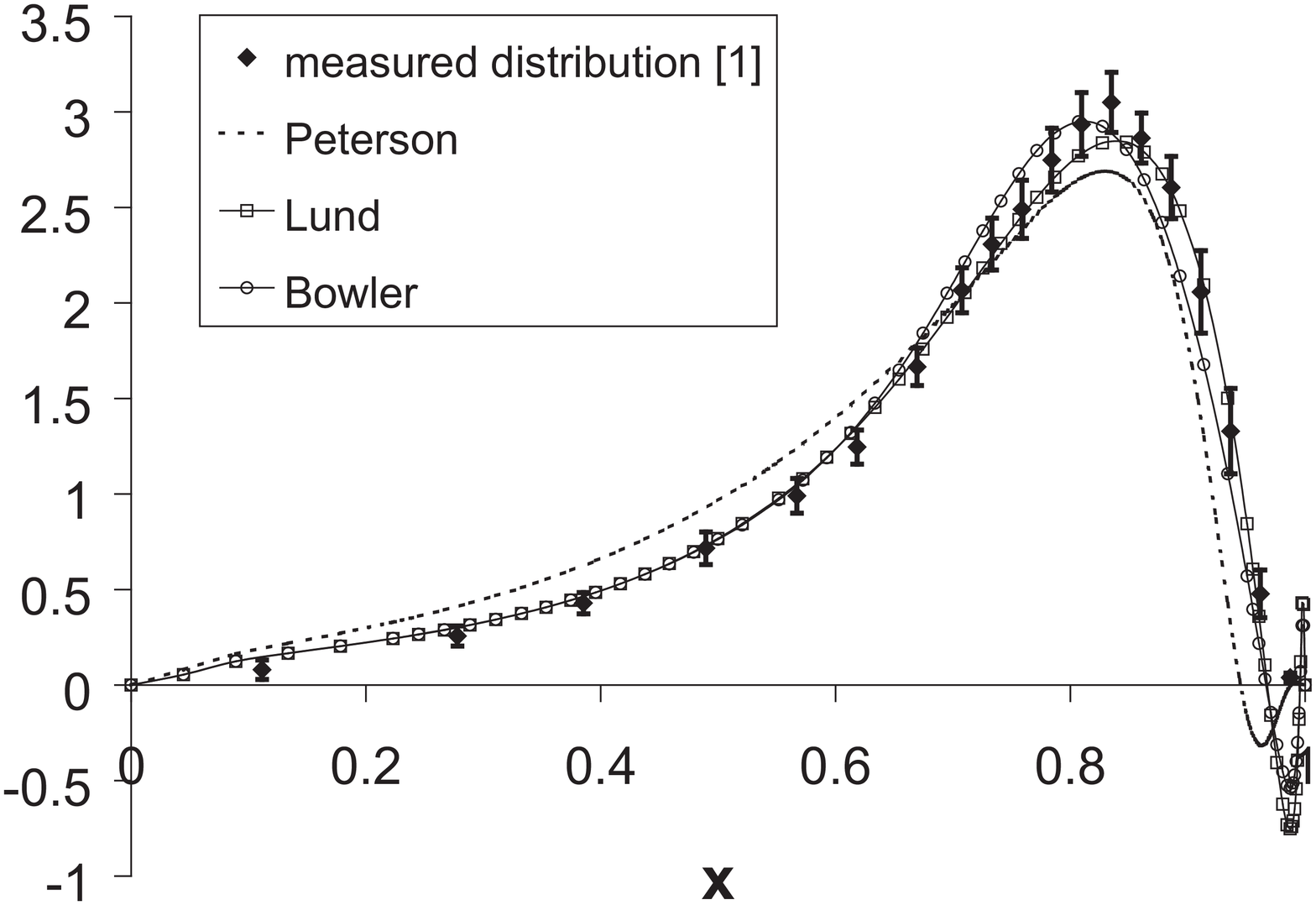, 
width=7cm,height=7cm}
          \epsfig{file=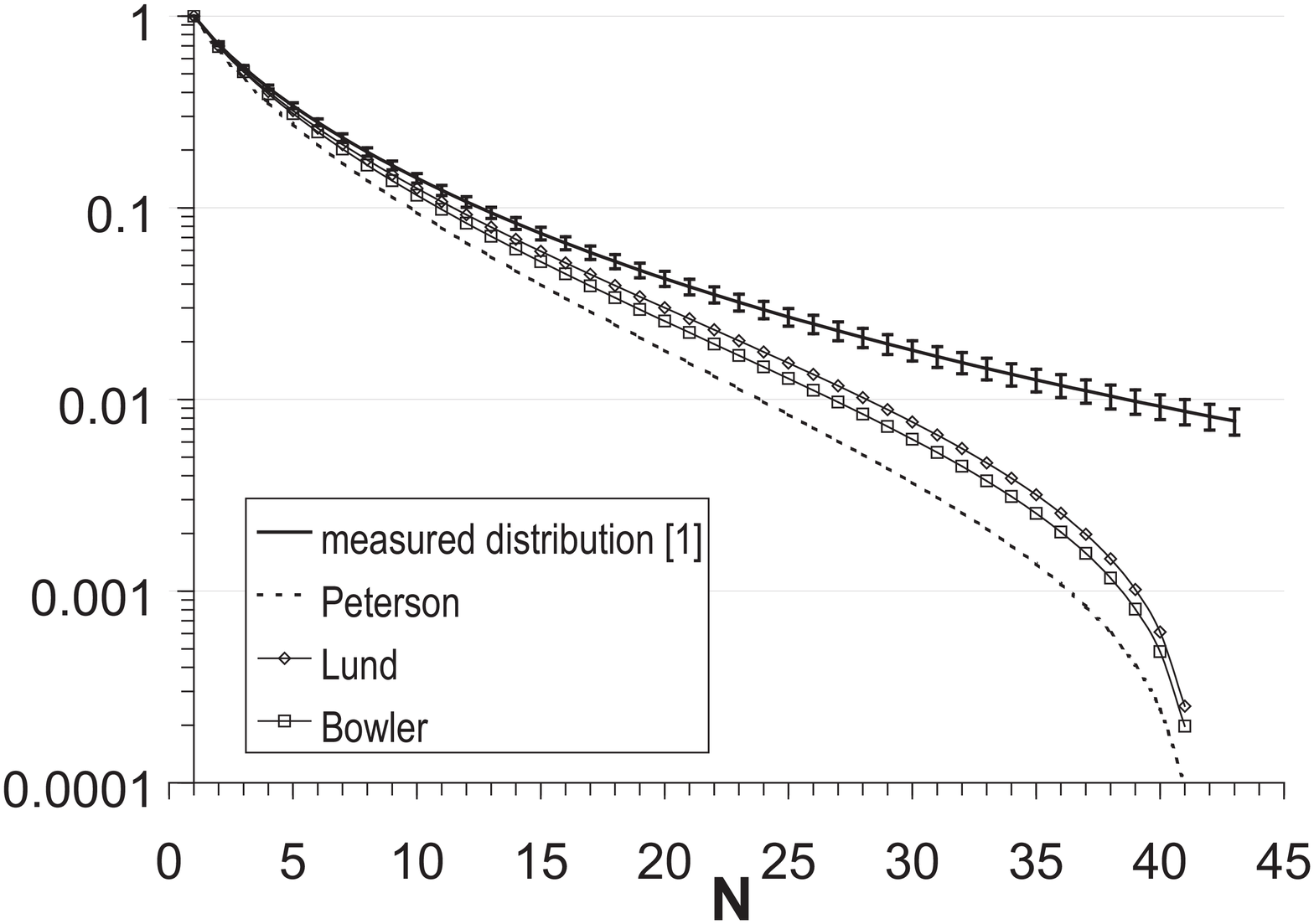,
width=7cm,height=7cm}}

  \end{center}
  \caption[]{\it {Comparison between the measured and fitted $x$-distributions
using different models.
Left: the measured binned-distribution in $x$ is compared with fitted results. 
Right: moments of the corresponding distributions are compared.}
   \label{fig:fold}}
\end{figure}

But, as these models have no contribution above $x=1$, their
folding with the perturbative QCD component cannot compensate
for the non-physical behaviour of the latter. 
The folded distribution is oscillating at large $x$-values 
(see Figure \ref{fig:fold}-Left).
In particular, the predicted
value in the last measured $x$-bin, which is found to be in reasonable
agreement with the measurements after the fitting procedure when using
the Lund and Bowler models, results from a large cancellation between
a negative and a positive contribution within that bin. This is more clearly
seen when considering the moments of the overall distribution, which
are given by Equation (\ref{eq:mellintb}). For moments of order $N$, the weight
$x^{N-1}$ introduces a variation within the $x$-bin size, which was
not accounted for in the previous fit in $x$, and effects are 
amplified mainly at large $N$ values
which correspond to the high $x$ region. 
This is illustrated in Figure \ref{fig:fold}-Right.

From this study, it results that all models have to be discarded
when folded with the NLL perturbative QCD fragmentation distribution.
The goodness of the fit in $x$, as measured by the corresponding $\chi^2$
value, does not
reflect all the information because it was not required that the folded
distribution remains physical (positive) over the $[0,1]$ interval.
This folding procedure has thus to be considered only as an exercise
and the non-perturbative QCD distribution has to be extracted from data.

\mysubsection{Proposal for a new parametrization}

As explained in Section \ref{sec:extract}, the non-perturbative QCD
component of the $b$-fragmentation distribution, has been extracted 
independently of any hadronic model assumption but it depends on 
the modelling of the perturbative QCD component. 

When a Monte-Carlo
generator is used to obtain the perturbative component, it can be verified
if the non-perturbative component has a physical behaviour for all $x$-values.
In Figure \ref{fig:jetset}, below $x=0.6$, $(4 \pm 3)\%$ of the integrated
distribution is negative. A larger deviation would have indicated some 
inconsistancy between
experimental measurements and gluon radiation, as implemented in
the generator. 

Such a test
cannot be made, a priori, when the perturbative QCD component is taken from an
analytic computation as this distribution is already unphysical in some 
regions.
The non-perturbative extracted distribution, as given in 
Figure \ref{fig:npx}, is precisely expected
to compensate for these effects. It can be noted that, also in this case,
the distribution is compatible with zero below $x=0.6$. This shows
that the perturbative QCD evaluation of hard gluon radiation is
in agreement with the measurements. The small spike, close to $x=0$,
related to the multiplicity problem \cite{ref:bass}
in the perturbative evaluation, has no numerical effect in practice.

To provide an analytic expression, which agrees better 
with the extracted point-to-point
non-perturbative QCD distribution, the following function
has been used:
\begin{eqnarray}
{\cal D}_{non-pert.}(x)&=&N({\bf p})
\exp{-\frac{(x-x_0)^2}{2 \sigma_-^2}}~{\rm for}~x<x_0;\nonumber \\
&=&N({\bf p})\exp{-\frac{(x-x_0)^2}{2 \sigma_+^2}}~{\rm for}~x_0<x<1;\nonumber \\
&=&N({\bf p}) x^{-N_0}~{\rm for}~x>1;~{\bf p}=(x_0,\sigma_-,\sigma_+)
\label{eq:param}
\end{eqnarray}
It corresponds, for $0<x<1$, to Gaussian distributions with different
standard deviations when $x$ is situated on either sides of $x_0$.
As explained in Section \ref{sec:extract}, the behaviour of the non-perturbative
distribution for $x>1$ is related to the presence of a zero in
${\cal D}_{pert.}(N)$ located at $N=N_0$. When the perturbative distribution
is taken from a Monte-Carlo, there is not such a zero and it has been
considered that ${\cal D}_{non-pert.}(x)=0$ when $x>1$.
When the perturbative distribution is taken from
theory, the value of $N_0$ has been fixed to 41.7.
The value for $N_0$ depends a lot on the central values for 
the parameters and the scales, adopted in the perturbative QCD evaluation
\footnote{$N_0$ is independent on the value assumed for the other scale
$\mu=\mu_F$.}:
\begin{equation}
N_0= 41.7^{-5.8}_{+4.5}(\Lambda^{(5)}_{QCD})\pm 2.5(m_b)
^{+34.6}_{-21.7}(\mu_0=\mu_{0F}) 
\label{eq:zero}
\end{equation}

$N({\bf p})$ is a normalisation factor such that the integral
of the expression, given in Equation (\ref{eq:param}),
between $0$ and $\infty$ is equal to unity.

The parameter values, obtained when the perturbative QCD component is 
taken either from JETSET or from theory, are given in Table \ref{tab:param}.
It can be noted that the shape of the non-perturbative QCD distribution
is similar in the two cases, the maximum being displaced to an higher value
in the latter.

\begin{table}[htb]
\begin{center}
  \begin{tabular}{|c|c|c|c|c|}
    \hline
 &$x_0$ &$\sigma_-$  & $\sigma_+$ & $\chi^2/NDF$\\
    \hline
JETSET &$ 0.911\pm 0.006 $ & $0.088\pm0.006  $ & $0.045\pm 0.005$ & $ 3.0/2  $\\
    \hline
NLL pert. QCD &$0.955\pm 0.011  $ & $0.087 \pm 0.006$ &$0.06^{+0.04}_{-0.02}$ & $ 0.7/2  $ \\
\hline
  \end{tabular}
  \caption[]{\it {Values for the fitted parameters of the non-perturbative QCD component corresponding to JETSET and NLL perturbative QCD using data from 
\cite{ref:aleph1}}.
  \label{tab:param}}
\end{center}
\end{table}

\begin{figure}[htb!]
  \begin{center}
    \mbox{\epsfig{file=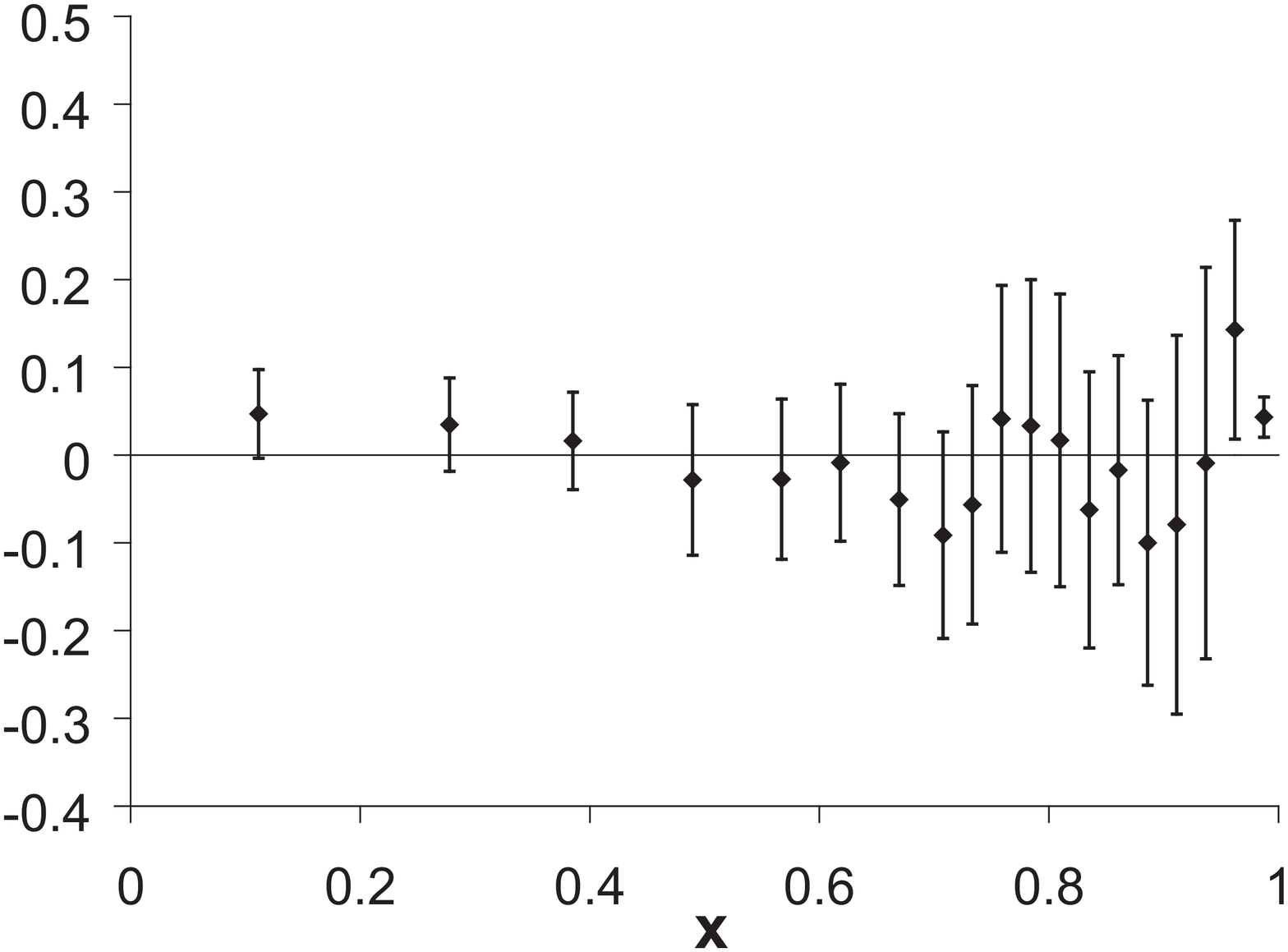, 
width=7cm,height=7cm}
          \epsfig{file=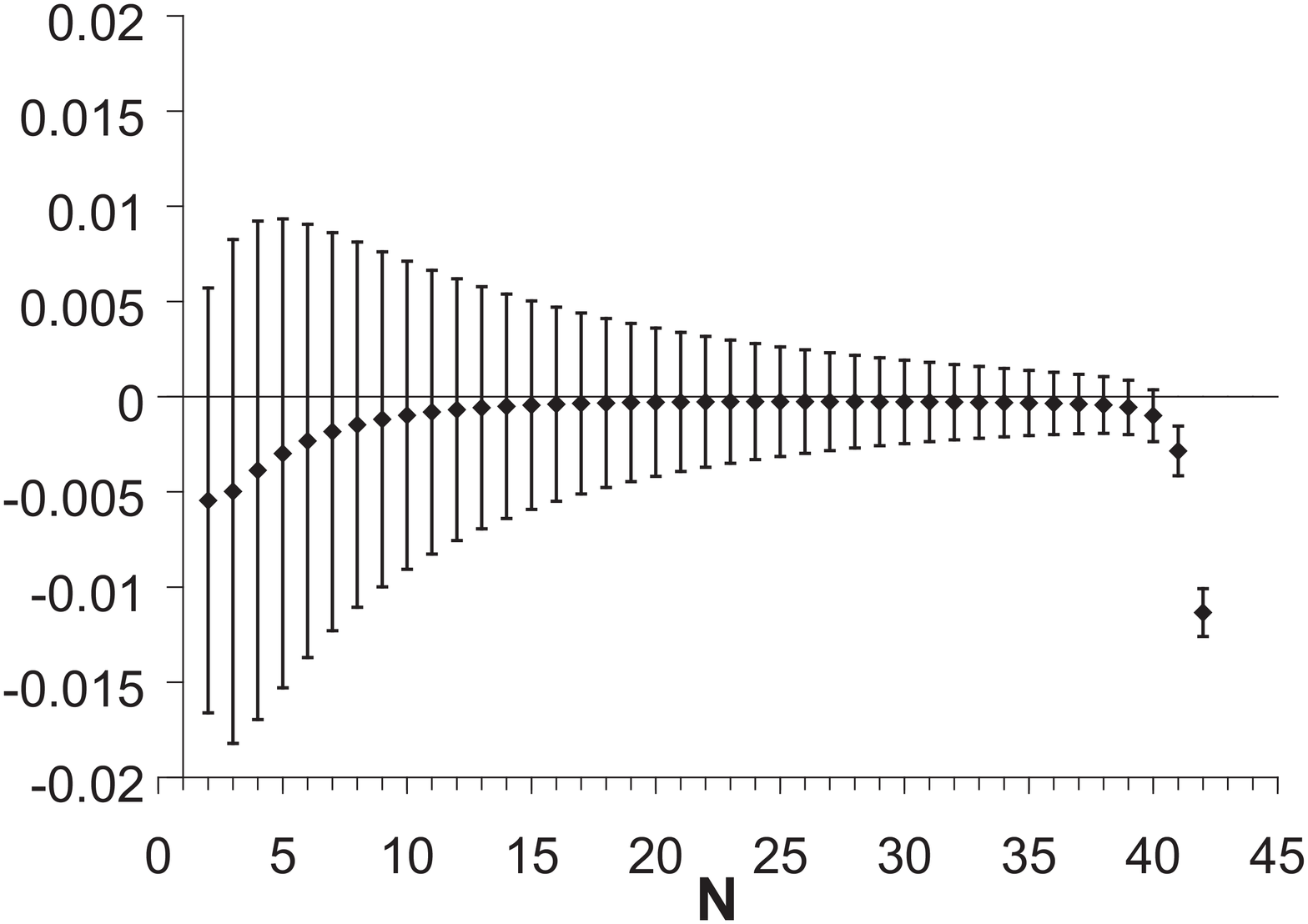,
width=7cm,height=7cm}}

  \end{center}
  \caption[]{\it {Comparison between the fitted and measured 
$b$-quark fragmentation distributions.
Left: Differences between the fitted and the measured distributions
in each $x$-bin are shown and the total error bars, relative to the measurements,
are displayed. The fitted results correspond to the 
averaged values obtained from
Equation (\ref{eq:start}) over the bin.
Right: Differences between the fitted and measured
moments of the $b$-fragmentation distribution.
Fitted moments result from the product given in Equation (\ref{eq:mellintb})
in which $\tilde{{\cal D}}_{non-pert.}(N)$ are the moments of the 
fitted non-perturbative QCD distribution
corresponding to Equation (\ref{eq:param}) and $\tilde{{\cal D}}_{pert.}(N)$
corresponds to the analytic computation \cite{ref:catani1}.}
   \label{fig:comp_fit}}
\end{figure}

    
Comparison, 
in $x$- and in moment-space, 
between the measured and fitted distributions 
are given in Figure \ref{fig:comp_fit}. For moments higher than 40, the 
fitted moments strongly deviate from the measurements as one is approaching
the Landau pole and the formalism of \cite{ref:catani1} is no longer valid. 

\mysection{Conclusions}
\label{sec:conclusion}

The measured $b$-quark fragmentation distribution has been analysed
in terms of its perturbative and non-perturbative QCD components.

The $x$-dependence of the fragmentation distribution has been extracted
in a way which is independent of any model for non-perturbative hadronic
physics. It depends closely on the way the perturbative QCD component has been evaluated. The obtained distribution differs markedly from those expected from various models.

Below $x=0.6$, this distribution is compatible with zero indicating that
most of gluon radiation is well accounted by the perturbative QCD component
evaluated using the LUND parton shower Monte-Carlo or computed analytically.

As the non-perturbative QCD distribution is evaluated for any given value 
of the $x$-variable it can be verified if it remains physical over the
interval $[0,1]$ when used with a Monte-carlo generator which provides
the perturbative component. The evidence for unphysical regions would
indicate that the simulation or the measurements are incorrect.
There is not such an evidence in the present analysis.

 Above $x=0.6$,
the obtained distribution is similar in shape with those expected
from the Lund symmetric \cite{ref:lund1} or Bowler \cite{ref:bowler1}
models, when the perturbative QCD component is taken from JETSET. 
This is no longer true, 
also for these two models, when the perturbative QCD component is taken from 
the analytic result of \cite{ref:catani1}. 
It has been found that,
because of the analytic behaviour of the perturbative QCD component,
the non-perturbative QCD distribution must be extended above $x=1$.
The $x$-behaviour of the non-perturbative component, for $x>1$, is
determined by the possible existence of a zero in
$\tilde{{\cal D}}_{pert.}(N)$, for $N>0$. 
When the perturbative component has non-physical aspects,
it is thus not justified to fold it with any given physical model.
An approach has been proposed to solve this problem and
a parametrization of the obtained distribution has been provided.

The non-perturbative component, extracted in this
way, is expected to be valid in a different environment than $\epem$ 
annihilation, as long as the perturbative QCD part is evaluated within
the same framework (analytic QCD computation or a given Monte Carlo generator), and using the same values for the parameters entering into this evaluation as $m_{b}^{pole}$, $\Lambda^{(5)}_{QCD}$ or generator tuned quantities.
It is planned to do a similar analysis using the improved analytic
computation of the perturbative QCD $b$-fragmentation distribution
given in \cite{ref:catani2}.


\mysection{Acknowledgements}
We thank M. Cacciari, S. Catani and M. Fontannaz for their help to understand
physical and technical aspects of the theory of $b$-quark fragmentation.
We thank also our colleagues from DELPHI, from the Institute for Experimental Nuclear Physics, Karlsruhe University, for fruitful discussions.

\noindent The work of E. Ben-Haim is supported by EEC RTN contract HPRN-CT-00292-2002.

\end{document}